# Soft Modes and Local Structural Transitions in Pb-free Ba(Ti$_{0.8}$Zr$_{0.2}$)O$_3$-$x$(Ba$_{0.7}$Ca$_{0.3}$)TiO$_3$ ($x$ = 0.5): Pressure- and Temperature-dependent Raman Studies

**Yu-Seong Seo and Jai Seok Ahn**[*]

*Department of Physics, Pusan National University, Busan 609-735, Republic of Korea*

**Il-Kyoung Jeong**

*Department of Physics Education, Pusan National University, Busan 609-735, Republic of Korea*

We report our Raman studies of a new lead-free relaxor ferroelectrics, Ba$_{0.85}$Ca$_{0.15}$Ti$_{0.9}$Zr$_{0.1}$O$_3$ (BCTZO). The Raman modes of BCTZO are compared with those of BaTi$_{0.8}$Zr$_{0.2}$O$_3$ (BTZO), and BaTiO$_3$ (BTO). Also, they are compared with the eigenmodes of BTO calculated by using an *ab-initio* quantum-mechanical frozen-phonon method. The sharp mode at 321 cm$^{-1}$ of BTO, reported as a coupled mode showing an interference effect, becomes progressively broader in BTZO and BCTZO. This behavior, together with a broadening of the 527-cm$^{-1}$ mode, suggests that the mode-coupling is weakened in BTZO and BCTZO. The structural transitions of BCTZO were investigated as functions of pressure at pressures below 20 GPa and of temperature at temperatures below 600 K. Three characteristic pressure-induced transitions, on each at 2.5, 5.0, and 13.0 GPa, were found. The transitions are suggested by the drastic changes in phonon modes (two softening modes, one each at ~ 300 and ~ 530 cm$^{-1}$) and by the transformation of the intensity profile. A temperature-induced transition was found at a Curie temperature of ~ 380 K, where the average structure changes from tetragonal to cubic. It is accompanied by a softening mode at ~ 530 cm$^{-1}$. The phonon spectrum of BCTZO suggests that its local environment is close to that of BTZO. However, the characteristic pressures of BCTZO are close to those of BTO. The sequence of pressure-induced transitions in both BCTZO and BTZO illustrate rich interplay between the long-range averaged structure and the short-range local order such that four distinguishable phases are suggested: tetragonal, locally ordered but compensated cubic, disordered cubic, and ideal cubic. We found that the critical pressures are plausibly related to the average crystal lattice.

* Email: jaisahn@pusan.ac.kr

 Fax: +82-51-513-7664 (Dept.)

# I. INTRODUCTION

Power-efficient high-performing electromotive devices, such as sensors, actuators and transducers, are needed in small electronics, industrial or medical applications [1, 2]. Such applications require a material with a large piezoelectric coefficient, $d_{33}$, and the most commercialized one is Pb(Zr,Ti)O$_3$ (PZT) [3]. As lead-based electronics increase environmental concerns, lead-free alternatives, (Bi,Nb)TiO$_3$-BaTiO$_3$, (Na,Bi)TiO$_3$, (Na,K)NbO$_3$, Ba(Ti,Zr)O$_3$, etc., have been developed [4]. Relaxor ferroelectrics have large electro-mechanical coupling strength, high dielectric breakdown strength, maximum achievable strain, and temperature insensitivity [5]. Also, relaxors have advantages as transducers for medical imaging because they provide large acoustic energy density and broadband capabilities. Recently, exceptionally large piezoelectric properties, with $d_{33}$ = 600 pC/N, were reported in Ba(Ti,Zr)O$_3$-$x$(Ba,Ca)TiO$_3$ (or BTZ-$x$BCT) ceramics by Liu and Ren [6]. The phase diagram of BTZ-$x$BCT showed three phases, rhombohedral, tetragonal and cubic, connected at a triple point, but not an in-between monoclinic/orthorhombic phase at the morphotropic phase boundary (MPB) as in PZT. The intermediate phase is usually believed to be indispensible for a polarization rotation mechanism across the MPB and for large $d_{33}$ [7]. Therefore, many questions remain to be answered regarding the superior piezo-properties of BTZ-$x$BCT. BTZ-$x$BCT is a relaxor-ferroelectric similar to PMN-$x$PT [8]. Therefore, its local structure, polar nano region, and dipolar domain-dynamics will be important in understanding its properties further. Neutron PDF analyses already provided notable local structural features departing from those of the given crystallographic structure, such as [111] off-centering of the Ti ion in the TiO$_6$ octahedron, a trigonal 3:3 Ti-O length distribution, and a single Zr-O bond-length in both BTZ and BTZ-$x$BCT [9-11]. A Raman spectroscopy is a promising technique to investigate relaxor ferroelectrics because it is a local probe suitable to the nanoscale physics of relaxors [12, 13]. Also, it provides phonon modes that deliver group symmetrical information on the crystal lattice through symmetry-coupled atomic motions. Its flexibility has been extensively utilized in bulk-phase transition researches [14-16]. In this manuscript, we report our Raman spectroscopic investigations on BTZ-$x$BCT ($x$ = 0.5). We applied pressure and changed temperature to induce structural transitions between the tetragonal, rhombohedral, and cubic phases. We found in-between phases, illustrating the rich interplay between the long-range average structure and the short-range local order.

## II. EXPERIMENTAL DETAILS

High-quality polycrystals, BaTi$_{0.8}$Zr$_{0.2}$O$_3$ (BTZO) and Ba$_{0.85}$Ca$_{0.15}$Ti$_{0.9}$Zr$_{0.1}$O$_3$ (BCTZO), were used for the present study. The samples were prepared by using a solid-state reaction method with stoichiometric amounts of BaCO$_3$, CaCO$_3$, TiO$_2$, and ZrO$_2$. The samples were sintered at 1723-1773 K several times, for a total of more than 68 hours in air, with intermediate grindings. The properties of the samples were carefully characterized through X-ray diffractometry, dielectric-constants measurements, and neutron diffractometry, as describedd elsewhere [10, 11].

The Raman spectra were obtained with standard confocal optics: a confocal sampling chamber, a monochromator ($f$ = 50 cm) equipped with a notch-filter, and a Peltier-cooled CCD camera. An ×50 objective was used both to focus and to collect the light in a back-scattering geometry. As an excitation source, the 488.0-nm line of an argon-ion laser was used. The laser power at the focal spot (2-3 $\mu$m in diameter) was kept at 12-15 mW by using a graded attenuation filter. A diamond-anvil pressure cell was used for high-pressure measurements. A mixture of 4:1 methanol-ethanol was used as a pressure medium. The applied pressure was varied up to 20 GPa (~ 200 kbar) and was calibrated *in-situ* by using ruby crystallites as a pressure standard [17]. A Linkam temperature-stage was employed for the temperature controlled experiments; temperatures ranged from 100 to 600 K, with a stability of 0.1 K.

## III. RESULTS AND DISCUSSION

Raman spectra of BCTZO and BTZO were measured at high pressures up to 20 GPa, as shown in Fig. 1(a). The spectra show significant change at two pressure values, 4 and 14 GPa, for BCTZO. The results can be classified into three different zones.

The low-pressure zone, $P < 4$ GPa, shows the representative characteristics of ferroelectric BTO-related oxides in tetragonal, orthorhombic, or rhombohedral phases. Several modes overlap below 400 $cm^{-1}$. Separated modes are found at 500 and 700 $cm^{-1}$.

The high-pressure zone, $P > 14$ GPa, shows the typical behavior of a paraelectric, ordered, cubic phase. BTO at high pressure has a paraelectric cubic phase similar to that of STO. Ti is located at the center (1/2, 1/2, 1/2) of the oxygen octahedron in STO with $Pm$-$3m$ symmetry; thus, Raman scattering becomes weak and broad at ambient pressure [18]. Becasuse first-order Raman scattering is forbidden in the $Pm$-$3m$ crystal symmetry, the broad features have been interpreted as being due to either strong second-order Raman scattering or disorder-induced first-order scattering associated with local deviations from cubic symmetry [18]. If the broad features are due to the local, static disorder, atoms will go back to their ideal positions at high pressure. Thus, the Raman-forbidden modes will disappear with increasing pressure so that the scattering intensity will rapidly decrease as shown in Fig. 1(b). The maximum intensities in the range of 150-400 $cm^{-1}$ are plotted.

Spectra of the in-between zone, $P = 4$-14 GPa, show intermediate results changing from a ferroelectric, tetragonal to a paraelectric, ordered, cubic phase. They show the typical feature of a first-order transition of an order-disorder type [19, 20]. An off-centering of the Ti ion appears in the given pressure range, and the average structure is known to be a paraelectric, disordered, cubic structure. Spectra show a continuous change from a ferroelectric to a paraelectric phase, mimicking a two-phase mixture. Such a continuous transformation is clearly identified from the color map in Fig. 1(c).

1. Mode Assignments for Raman Spectra at Ambient Pressure

Figure 2 shows the Raman spectrum of BCTZO compared with those of BTZO and BTO measured at ambient conditions. The Raman spectrum of BCTZO resembles those of BTZO and BTO. The basic features of BCTZO or BTZO share close similarities to those of BTO, such as broad modes below 400 cm$^{-1}$ and narrow lines at 500 cm$^{-1}$ and around 700 cm$^{-1}$. If this is to be understood further, quantitative information is needed.

The scattering intensity profile, $I(\omega)$, was analyzed using a sum of Lorentzian functions as follows:

$$I(\omega) = \sum_i \frac{S_i \gamma_i^2}{(\omega - \omega_i)^2 + \gamma_i^2},$$

where $\omega_i$, $\gamma_i$, and $S_i$ correspond to the frequency, damping, and strength of the $i$-th mode. After least-squares refinements, we obtained mode parameters for each spectrum. The spectrum of BTO was analyzed with eleven modes, 133, 155, 173, 208, 266, 306, 347, 512, 539, 604, and 717 cm$^{-1}$. We needed eight modes for BTZO: 169, 238, 291, 514, 557, 626, 738, and 795 cm$^{-1}$; ten modes were needed for BCTZO: 157, 213, 258, 299, 523, 555, 644, 731, 766, and 811 cm$^{-1}$.

Theoretical values of Raman mode frequencies for BCTZO/BTZO are not available due to their structural complexity. Thus, we compare the modes in the Raman spectra with the calculated modes of BTO. For tetragonal BTO, phonon modes are decomposed into eight branches at the $\Gamma$-point: $\Gamma = 4E + 3A_1 + B_1$ [21]. The $E$ and the $A_1$ modes are both infrared (IR) and Raman active, and the $B_1$ is a Raman only mode. We obtained phonon frequencies by using a frozen-phonon method, which is implemented in quantum-mechanical *ab-initio* Hartree-Fock calculation software, CRYSTAL09 [22], utilizing Gaussian-type orbital bases. The second derivatives of the total energy were computed with respect to a finite atomic displacement. The phonon frequencies were calculated as 144($E$), 177($A_1$), 195($E$), 321($B_1$), 358($E$), 447($A_1$), 527($E$), and 653 cm$^{-1}$($A_1$). Here, we only tabulate frequencies for Both-active (IR and Raman) transverse optic (TO) phonon modes and a Raman-only mode. Compared with previous results [23], our calculation gave all real frequencies for tetragonal BTO. We also found the mode at the lowest frequency to severely depend on the choice of exchange-correlation functionals among several density functional approximations, such as local density approximation (LDA), generalized gradient approximation (GGA), B3LYP, Hartree-Fock (HF), etc. Except the first mode, all other modes were reliably reproducible irrespective of the choice of the functionals. Eigenvectors corresponding to well-separated modes, 321 ($B_1$), 527 ($E$), and 653 cm$^{-1}$ ($A_1$), are indicated as characteristic atomic motions in the inset of Fig. 2. In the 321 ($B_1$) cm$^{-1}$ mode, planar oxygen atoms around the Ti ion perform a buckling motion with respect to Ti ion. In the 527 ($E$) cm$^{-1}$ mode, the TiO$_6$ octahedron simultaneously shows a distortion in the TiO$_4$ plane and a bending of the Ti ion and apical oxygen atoms. The 653 cm$^{-1}$ ($A_1$) mode corresponds to an oscillatory stretching motion between the TiO$_4$ plane and apical oxygen atoms.

The most striking feature below 400 cm$^{-1}$ in the BTO spectrum is the sharp peak at 306 cm$^{-1}$. The narrow mode at 306 cm$^{-1}$ of BTO has been interpreted as a coupled mode showing an interference effect [24]. Such a sharp feature turns into a broad peak in BTZO. Further, it is less pronounced in BCTZO. The broadening of the 306-cm$^{-1}$ mode, which is coupled to the 512-cm$^{-1}$ mode as will be discussed below, suggests that the mode-coupling is progressively weakened in the BTZO/BCTZO system. In the broad structure below 400 cm$^{-1}$, the low-energy part at 150 cm$^{-1}$ is more pronounced in BTZO. In BCTZO, the broad structure consists of four overlapping peaks, as shown in Fig. 2

Features at 450-700 cm$^{-1}$ are decomposed into three Lorentzians, which are centered at 512, 539, and 604 cm$^{-1}$ in BTO. According to Sood et al. [24], the first narrow peak at 512 cm$^{-1}$ in BTO spectrum has been assigned as the $A_1$ coupled-oscillator mode (at 470 cm$^{-1}$ if uncoupled) bound with two other $A_1$ modes (at 190 and 320 cm$^{-1}$, if there is no coupling) based on the classical, coupled-oscillator model. The peak width becomes much larger in BTZO/BCTZO, similar to the 306 cm$^{-1}$ mode. Farhi et al. also noted that the mode-coupling becomes weak due to disorder introduced by Zr-substitution [25]. Furthermore, our phonon calculation indicates there is no $A_1$-type mode in close proximity of 500 cm$^{-1}$: $A_1$ modes are expected at 177, 447, and 653 cm$^{-1}$. Instead, an $E$-type mode is expected at 527 cm$^{-1}$.

Therefore, the three modes at 450-700 cm$^{-1}$ can be interpreted as $E$, $A_1$, and longitudinal optic (LO) modes.

The 717-cm$^{-1}$ mode in BTO has been reported as a characteristic LO mode [19, 21]. Note that our calculation predicted two LO modes, 739 ($E$-LO) and 852 cm$^{-1}$ ($A_1$-LO), of tetragonal BTO near 700 cm$^{-1}$. The 717-cm$^{-1}$ mode decomposes into two peaks in BTZO/BCTZO. The peak separation has been clearly evidenced in Zr-rich BTZO [25, 26], and the second peak is understood to be a characteristic feature of relaxors. It is puzzling if the high frequency component originates from Zr because it is heavier than Ti. A sum-frequency mode may be an alternative explanation; for example, the 717 cm$^{-1}$ is close to the sum of 512 and 208 cm$^{-1}$ in BTO. Likewise, the two modes, 738 and 795 cm$^{-1}$, in BTZO can be approximated by adding 514 cm$^{-1}$ to 238 or 291 cm$^{-1}$. Moreover, three modes (731, 766, and 811 cm$^{-1}$) are observed in BCTZO, as shown in Fig. 2: they can be obtained by adding 213, 258, or 299 cm$^{-1}$ to 523 cm$^{-1}$.

## 2. Pressure Dependence of Raman Modes

The pressure dependences of the individual mode frequencies are presented in Fig. 3 for BCTZO and BTZO. If we look closely into the evolution of the modes, we can find hints of structural transitions. For example, in BTZO, new modes develop at pressures above 5.6 GPa at 400 and 650 cm$^{-1}$, while all other modes remain. These indicate a transition to a different symmetry. Kreisel et al. also reported similar spectral signatures of a new (at least local) symmetry at pressures from 5.7 GPa in BTZO35 [27].

Based on the appearance and disappearance of modes, we can catalog a series of transitions in BCTZO. A new mode emerges at 400 cm$^{-1}$ at pressures above 2.5 GPa, but the 320-cm$^{-1}$ mode declines. At pressures beyond 5.0 GPa, the 500-cm$^{-1}$ mode disappears, and its spectral weight moves to nearby, ~ 400 and ~ 600 cm$^{-1}$, regions. Also, modes at 100-150 cm$^{-1}$ merge at pressures above 5.0 GPa. At pressures above 13 GPa, the 300-cm$^{-1}$ mode becomes prominent and the peak at ~ 800 cm$^{-1}$ (pressure-shifted from 700 cm$^{-1}$) becomes more noticeable. Also, the 500-cm$^{-1}$ mode resurfaces at pressures above 13 GPa. With the same method, the transition-pressures of BTZO are obtained as 5.6, 8.9, and 15 GPa. Note that only two transition-pressures, 5.7 and 15.1 GPa, were previously reported for BTZO35 [27], but another transition-pressure at 8.9 GPa could be inferred from the diminishing 500-cm$^{-1}$ mode in the data of Ref. 27. In the case of BTO, transitions have been reported at 2.5, 5.2, and 10 GPa [19, 28]. The series of transitions is accompanied by structural-property changes as already reported by many other groups. Let us index the transition-pressures as $P_1$, $P_2$, and $P_3$, in increasing order.

Let us discuss the continuous, monotonic changes of the mode frequency as a function of pressure, as demonstrated in Fig. 3. Most modes, except a few soft modes, blue-shift monotonically. Especially, the high-frequency modes above 700-cm$^{-1}$ move to higher frequencies with increasing pressure at a constant rate of ~ 5.1 cm$^{-1}$/GPa, for both BTZO and BCTZO. The pressure-rates of the frequency curves for the ~ 720 cm$^{-1}$ peaks are found as 4.4 cm$^{-1}$/GPa ($P$ = 0-19.2 GPa) for BTZO, and 4.8 ($P$ < 5 GPa) or 5.6 cm$^{-1}$/GPa ($P$ = 0-19.4 GPa) for BCTZO. These values are very close to the reported values, 4.5 cm$^{-1}$/GPa ($P$ = 2-5 GPa) for BTO and 4-5 cm$^{-1}$/GPa ($P$ = 0-27 GPa) for BTZ35 [19, 27]. This pressure dependence is explained with the mode-Grüneisen effect [29], which relates the blue-shift to volume shrinkage. In the high-pressure range above $P_1$, the crystal system is assumed to be cubic; thus, the 720-cm$^{-1}$ mode of BTO/BTZO/BCTZO can be compared with 600-cm$^{-1}$ mode of SrTiO$_3$. Interestingly, the rate of change of the frequency with increasing pressure is obtained as 9.0 cm$^{-1}$/GPa ($P$ = 0-16 GPa) for SrTiO$_3$ [18].

Tetragonal symmetry is broken at $P_1$; thus, the ferroelectric phase changes into a paraelectric, compensated, cubic phase. Here, the phase has local polar blocks with off-centered Ti within the oxygen octahedral cage, but the polarities of the blocks are mutually cancelled by non-polar atomic arrangements on a larger scale. Thus, the average structure is cubic. Such behavior is commonly observed in relaxor ferroelectrics, and often it has been called a polar nano region, when the size is large enough to form domain structures on a nanometer scale. On approaching $P_1$, the 300-cm$^{-1}$ mode ($B_1$) of BTZO begins to soften and damp, as shown in Fig. 4. Frequency-softening has been found in BTO and BTZO35 [19, 27]. BCTZO also shows mode-softening and increased damping at pressures below $P_1$ (2.5 GPa). However, they are less pronounced than in BTZO or BTO. The reason is that the mode overlaps with nearby modes, as shown in Fig. 2. Thus, the refinement is more prone to error, as indicated with the large error bars in Fig. 4.

No structural information related to a transition at a pressure of $P_2$ has been reported. The only similar information is the critical pressure of BTO, ~ 6.5-8.4 GPa [30, 31]. Zhong et al. predicted a quantum phase transition from a tetragonal (orthorhombic or rhombohedral) to a cubic phase at $P \approx 8.4$ GPa [30]. Ishidate et al. observed a classical- to quantum-limit crossover at $P \approx 6.5$ GPa by using the critical exponent analysis on the dielectric constant [31]. The transition at a pressure of $P_2$ is clearly demonstrated as a softening behavior of the ~ 500-cm$^{-1}$ ($E$) mode, as shown in Fig. 4. The mode becomes soft, and damping $\gamma$ increases towards the transition: the critical pressure $P_2$ is measured as 6 GPa for BCTZO and 9 GPa for BTZO. Above that pressure, the average structure is known to be a cubic. However, it is not the same highly-symmetric cubic structure as in SrTiO$_3$, where Ti is located at the center of the oxygen octahedron, as was discussed above.

There are two different pressure regions at pressures above $P_2$. If we closely look at Figs. 1(a)-(c) at pressures between $P_2$ and $P_3$, we can see a continuous change in the spectra from $P_2$ to $P_3$. This resembles the typical observation of inter-mixing between two phases. Structural refinements are highly anticipated, but at this stage, a disordered (locally) cubic phase may be a proper collocation between the compensated cubic and the highly-symmetric one. Spectra at pressures above $P_3$ show the typical behavior of a highly-symmetric cubic phase, featuring a weak scattering intensity and two dominant bands, ~ 300 and ~ 800 cm$^{-1}$, similar to SrTiO$_3$ [18].

## 3. Temperature Dependence of Raman Spectra of BCTZO

BCTZO shows a series of structural transitions in a sequence of rhombohedral(*R*)-tetragonal(*T*)-cubic(*C*) from zero temperature onwards: a *R-T* transition at 287 K and a *T-C* transition at 366 K [21]. Figure 5(a) summarizes the Raman spectra of BCTZO at various temperatures between 100 and 600 K. Nine modes (above 100 cm$^{-1}$) are clearly resolved at 100 K. The spectrum changes gradually with increasing temperature. Modes become simplified and overly-damped at 600 K. However, no notable feature was observed near the *R-T* transition temperature. This observation is clearly different from the series of transitions reported in BTO [32]. This result partly agrees with the recent neutron pair-distribution-function analysis on BCTZO by Jeong and Ahn: the average structure varied with temperature, but the local structure remained nearly unaffected at the *R-T* crossover [11].

The *T-C* transition in BCTZO is accompanied by characteristic changes in phonon modes. The ~ 300-cm$^{-1}$ mode is recognizable at temperatures up to 350 K, but is buried in one broad band at temperatures above 400 K, as shown in Fig. 5(a). An analogous mode-damping tendency is observed at pressures above $P_1$, the threshold between the tetragonal and the compensated cubic phases, in the pressure-Raman result shown in Fig. 4. Figure 5(b) shows the frequency and the damping of the 500-cm$^{-1}$ (*E*) mode as functions temperature. The *E* mode shows mode-softening when approaching 375 K from below, and its damping increases rapidly at temperatures above 350 K. The *E* mode shows a similar behavior as a function of pressure above $P_2$, characterizing a transition from a compensated cubic to a disordered cubic phase, as was discussed above. The two transitions, one each at $P_1$ and $P_2$, therefore, seem to converge into a single transition in temperature within a 50-K interval.

## 4. On the Similarity of BCTZO with BTZO or BTO

Let us discuss our puzzling observations in BCTZO compared to those in BTZO and BTO. The Raman modes of BCTZO are similar to those of BTZO. However, the transition pressures of BCTZO are close to BTO ones. The similarity of Raman modes between BCTZO and BTZO suggests that their local structures are analogous too: they are composed of a $ZrO_6$ block with Zr on its center and a $TiO_6$ octahedron with a [111]-off-centered Ti ion [9-11]. The issue of the critical pressures is plausibly related to the average structures. BCTZO is obtained by substituting ions in BTO: Ca at the *A*-site and Zr at the *B*-site. BTZO has a larger cell size than BTO because Zr is bigger than Ti. However, Ca compensates for such an effect in BCTZO, for Ca is smaller than Ba. The determined cell size of BCTZO, 64.3 Å$^3$, is very close to the 64.4 Å$^3$ for BTO at ambient conditions [11, 33]. This implies that the transition pressures are closely correlated with the cellsizes, which provides a hint to the solution to the puzzle of only 10% zirconium being required to get a relaxor phase in the BTZ-*x*BCT system, with 28% being required in the BTZ case [6].

## IV. CONCLUSION

We showed using Raman spectroscopy and *ab-initio* phonon calculations that the pressure dependence of Raman spectra of BCTZO/BTZO are characterized by a series of transitions governed by a competition between the long-range averaged structure and the short-range local order. The Raman spectrum at ambient pressure of BCTZO is similar to that of BTZO, but different from that of BTO in that mode-coupling is weakened and high-frequency modes are decomposed. The pressure dependence shows the complex characteristics of intermediate phases between ferroelectric tetragonal and paraelectric ordered cubic. An analysis of the pressure-dependent Raman modes shows that three structural transitions occur, mostly in accord with the previous reports. Temperature-insensitive Raman spectra agree with the recent neutron PDF analysis, implying that the local structure remains almost the same at the *R-T* transition [11].

# ACKNOWLEDGMENTS

This work is supported by the National Research Foundation of Korea (NRF) grant funded by the Ministry of Education, Science and Technology (MEST), No. 2012006641. The computation is supported by the Korea Institute of Science and Technology Information (KISTI) Supercomputing Center through contract No. KSC-2012-C2-36. IKJ was supported by the NRF grant funded by the MEST, No. 2012-0000345.

Figure Captions.

Fig. 1. Raman spectra of BCTZO at room temperature at applied pressures of $P$ = 0-20 GPa: (a) Raman scattering spectra at each pressure value, (b) maximum intensities of the 150-400 cm$^{-1}$ band: ●(BCTZO) and ▲(BTZO), and (c) intensity plots of BCTZO-spectra, background-subtracted and normalized to unity.

Fig. 2. Raman spectra of BTO (top), BTZO (middle), and BCTZO (bottom) at ambient conditions. Experimental data are drawn with solid lines. Best-fit curves are overlaid with dashed lines. Lorentzian decompositions are shown with dashed curves. BTO in 11 modes: 133, 155, 173, 208, 266, 306, 347, 512, 539, 604, and 717 cm$^{-1}$; BTZO in 8 modes: 169, 238, 291, 514, 557, 626, 738, and 795 cm$^{-1}$; BCTZO in 10 modes: 157, 213, 258, 299, 523, 555, 644, 731, 766, and 811 cm$^{-1}$. Characteristic atomic movements of the TiO$_6$ octahedron are shown as insets for the corresponding eigenmodes, 321 ($B_1$), 527 ($E$), and 653 cm$^{-1}$ ($A_1$), calculated for BTO.

Fig. 3. Pressure dependence of the Raman mode frequencies, $\omega$: (upper panel) BCTZO and (lower panel) BTZO. Vertical lines indicate the three transition pressures, $P_1$, $P_2$, and $P_3$, in increasing order.

Fig. 4. Pressure dependence of the $E$ mode (left panels: at ~ 500 cm$^{-1}$) and the $B_1$ mode (right panels: at ~ 300 cm$^{-1}$). For BTZO (▲) and BCTZO (●). Upper: phonon frequency, $\omega$; Lower: damping, $\gamma$. Lines are guides for the eye.

Fig. 5. (a) Temperature dependence of the Raman spectra of BCTZO for $T = 100$-$600$ K in 50-K steps. (b) Temperature dependence of the $E$ mode: mode frequency $\omega$ and damping $\gamma$. The vertical line indicates the ferroelectric transition temperature.

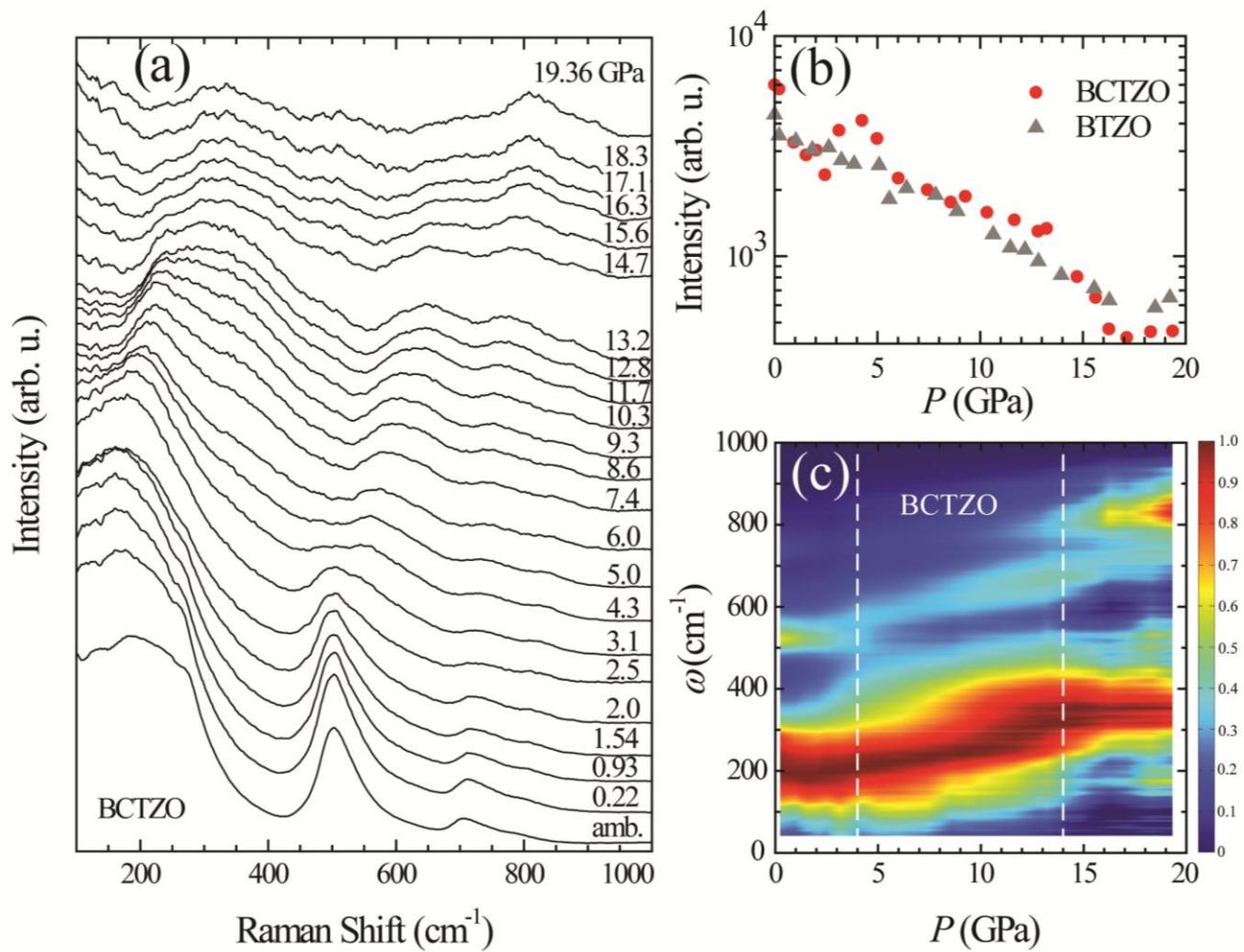

Figure 1 of Seo et al.

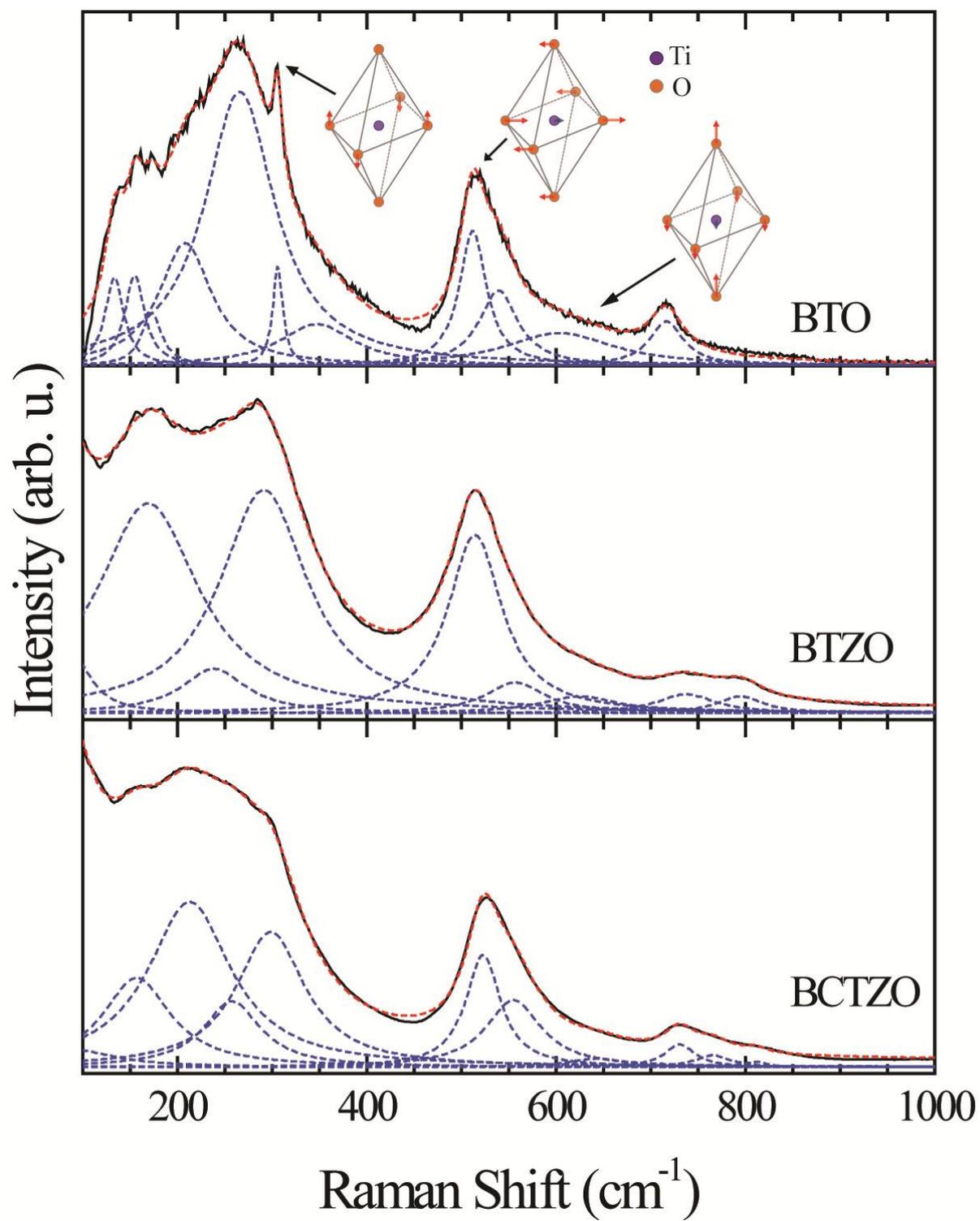

Figure 2 of Seo et al.

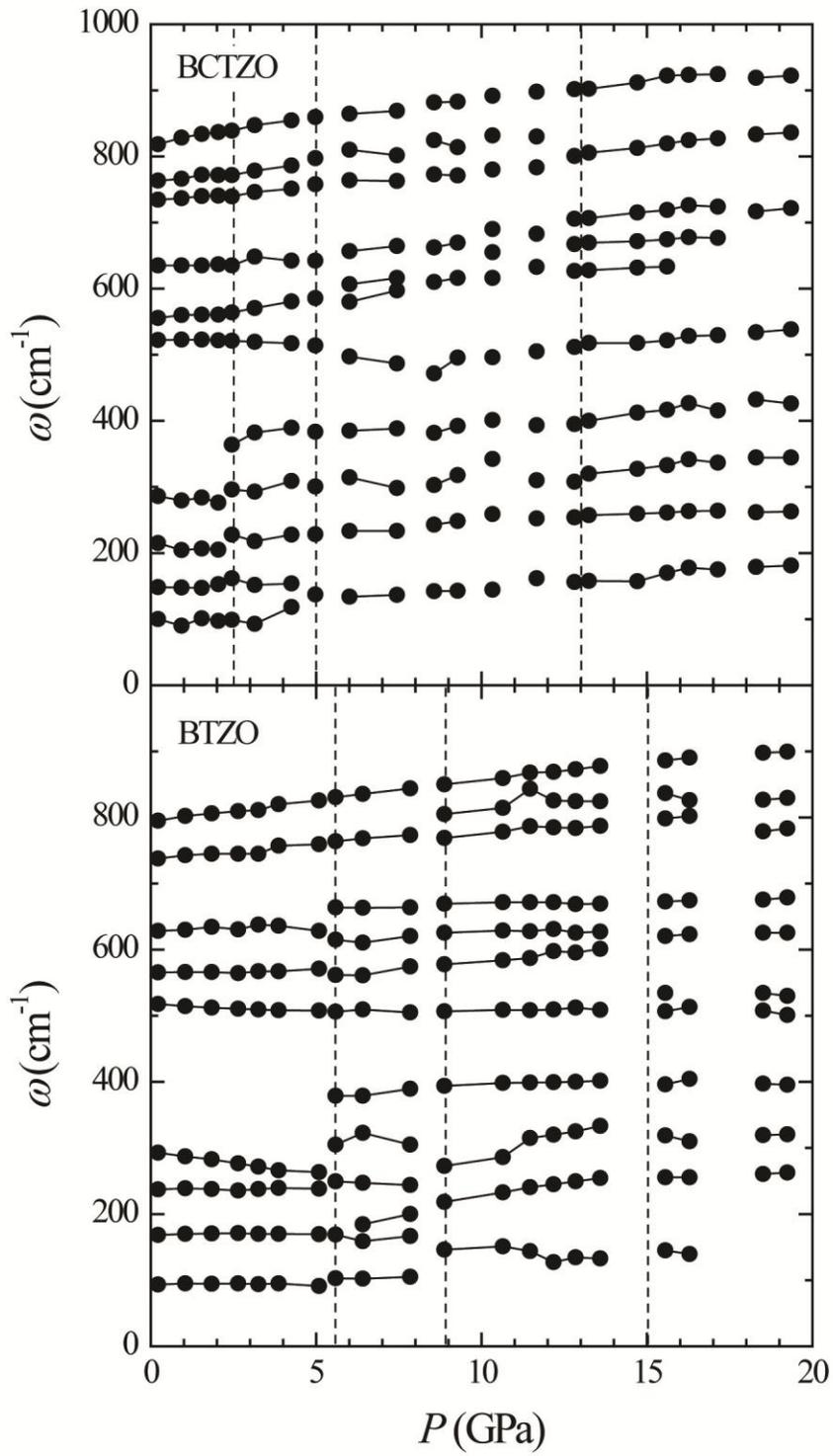

Figure 3 of Seo et al.

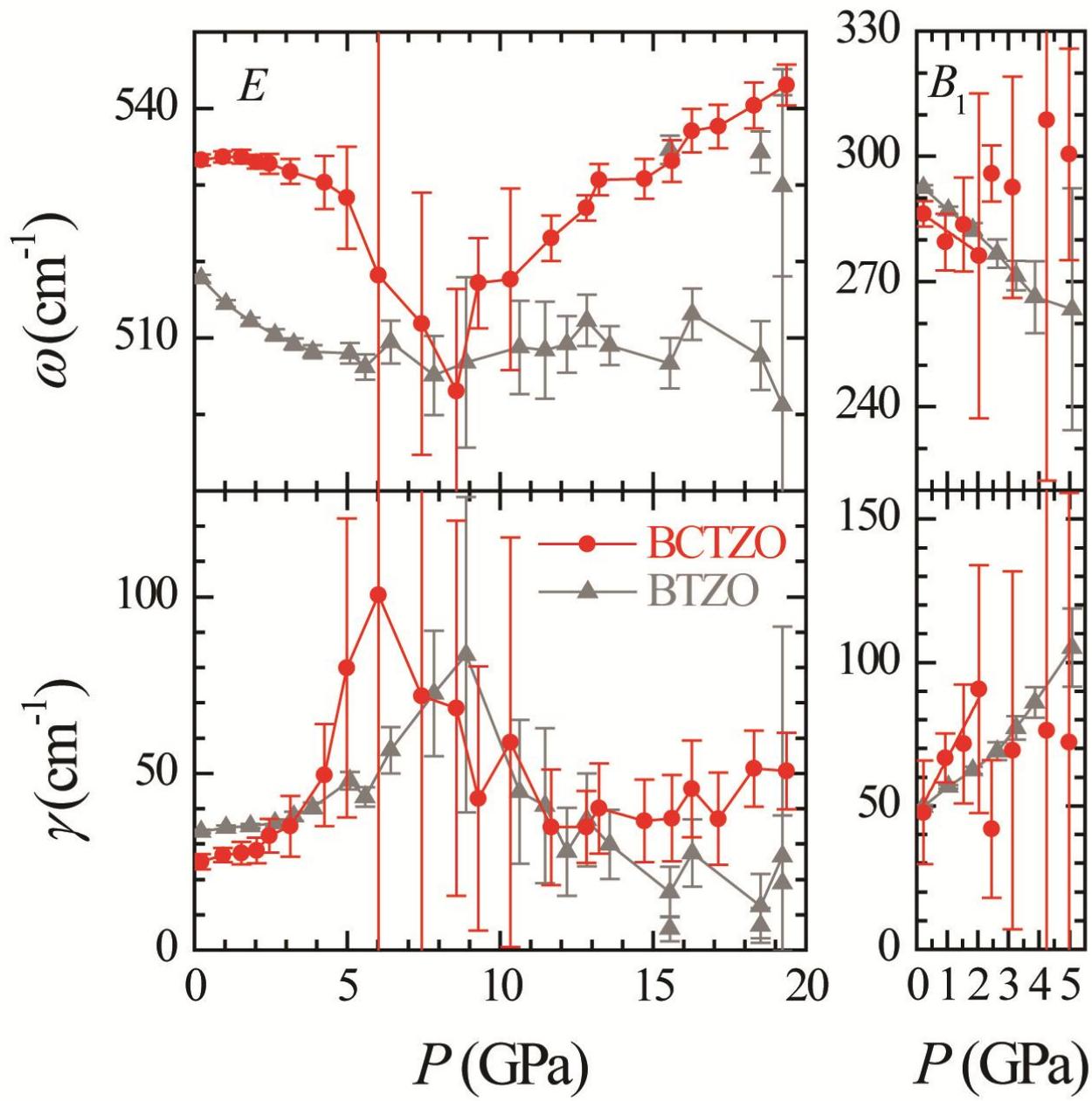

Figure 4 of Seo et al.

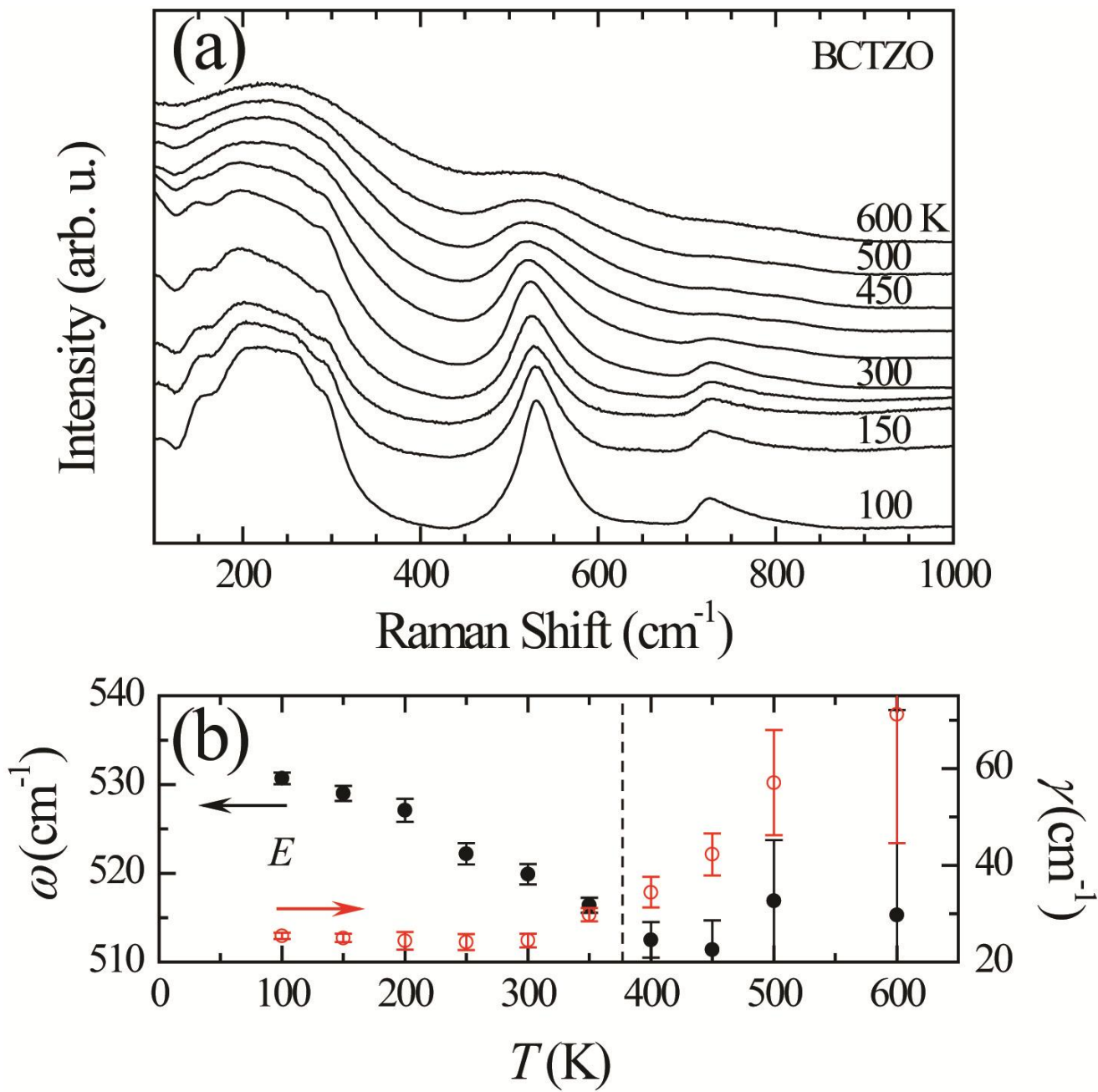

Figure 5 of Seo et al.